\documentclass[prb,floats,twocolumn]{revtex4}
\input epsf
\usepackage{graphicx}

\newcommand\beq{\begin{equation}}
\newcommand\bear{\begin{eqnarray}}
\newcommand\eeq{\end{equation}}
\newcommand\eear{\end{eqnarray}}
\usepackage{amsmath}
\begin{document}
\setlength{\topmargin}{0.0 cm}

\title[not]{\bf Soft Modes at the Stacking Faults in SiC Crystals} 

\author{Tiju Thomas$^1$,  Dhananjai Pandey$^2$ and Umesh V. Waghmare$^1$\footnote{Electronic mail: tijuthomas@jncasr.ac.in}} 
\affiliation{$^1$ T. S. U, Jawaharlal Nehru Centre for
Advanced Scientific Research,\\ Jakkur Campus, Bangalore 560 064, India.\\
$^2$ School of Materials Science and Technology, Banaras Hindu University, Varanasi - 221 005.}

\date{\today}
\begin{abstract}   
We use first-principles calculations based on density functional theory
to determine and understand the driving force of the observed stacking fault 
expansion in SiC. We verify the suggestion based on recent experiments
that the free energy difference between the faulted and the perfect structures
is responsible for this expansion, and show that its origin lies in a large
entropy associated with soft vibrational modes of the faulted SiC structure. 
These soft modes involve shearing of SiC on a long length-scale and are absent 
in related group IV semiconductors, such as Si, Ge and C.

\end{abstract}
\pacs{
}
\maketitle
Among the group IV semiconductors (Si, Ge, C, SiC), SiC is unique in
exhibiting hundreds of polytypes that differ only in the stacking sequence of
Si-C double layers \cite{VermaKrishna}. Due to its excellent physical
properties such as high hardness, low density and thermal expansion,
it has found numerous applications in automotive components, seal faces, armour,
mechanical seals, bearings, heat exchanger tubes, etc.
\cite{Perle}
As it is available in large single crystal wafer form\cite{Sudarshan}, 
SiC is also interesting as a wide band-gap semiconductor for electronic
applications: it is a promising material for the next generation 
power diodes owing to its low on-resistance, high break-down voltage, 
radiation stability and low switching losses\cite{Burak}. 
Recently, there have been efforts aimed at developing  
dilute magnetic semiconductors (DMS)\cite{Shaposhnikov} based on SiC
for use in spintronic devices. 

However, formation of line defects\cite{Konstantinov, Lendenmann}
 and expansion of stacking faults\cite{MKDas} in
SiC crystals are known to be possible causes of 
degradation of the performance of SiC-based power 
diodes. Understanding the mechanism of stacking fault expansion 
in SiC is important to its use in electronic devices and it may also shed light
on a related fundamental question of the stability of polytypes of SiC\cite{Pandey}. It is also
interesting to find out if this mechanism operates in other group IV semiconductors.

In a power electronic device current densities as 
high as 50-100 A/$cm^{2}$ lead to an enormous amount of heating 
and operational temperatures are higher 
than the normal ones. Optical emission microscopy (OEM) 
based experiments on 4H-SiC used in high current density 
conditions revealed stacking fault expansion in the 
basal plane of the system, establishing that hexagonal polytypes of SiC are
unstable at high temperatures with respect to formation and expansion of 
stacking faults\cite{MKDas}. Based on this work, free 
energy differences was suggested to be the driving force of 
stacking fault expansion.

Many groups have carried out first-principles 
investigation of the stability of various
polytypes of SiC and their stacking faults\cite{Lindefelt,Hong,Maeda}.
With very small energy differences involved, the exact ordering
energies of polytypes is a bit scattered in these studies. However, they all 
estimated very small but positive stacking fault energies. It is not clear 
yet what drives the stacking fault expansion in SiC at finite temperature.

In this work, we use first-principles calculations to determine free energy 
differences between different polytypes and stacking-faulted structures of
SiC through determination of configurational and vibrational entropies. 
Configurational and vibrational contributions to free energy are determined 
from the knowledge of full stacking fault energy surface \cite{Vitek}
and of the vibrational spectrum respectively. We show that soft shear modes
developed in the faulted structures drive their stability at finite temperature.

Diamond (or zincblend, labeled here as $3C$) structure, the common 
polytype of all group IV semiconductors, can be described 
as two interpenetrating FCC lattices separated by $[111]a/4$, $a$ being 
the lattice constant. Along [111] direction, triangular lattices of 
atoms described with high symmetry positions A, B and C are 
stacked periodically as (AaBbCc), where upper and lower case letters 
label the two FCC sublattices. In case of SiC,
carbon atoms occupy sites of the second sublattice. Atoms in the 
polytypes obtained with a variation in this stacking sequence 
(studied in this work) are all tetrahedrally coordinated. In the Hagg 
notation\cite{Hagg} used commonly in the representation of 
stacking sequences, (AaBb), (BbCc) and (CcAa) are denoted as (+), 
and the anti-cyclic stacking sequences eg. (BbAa) 
are denoted as ($-$). Thus, 3C-SiC (with AaBbCc stacking), 
4H-SiC (with AaBbAaCc stacking) and 6H-SiC (with AaBbCcAaCcBb stacking) 
polytypes are represented with periodic units of ($+++$), 
($++--$) and ($+++---$), respectively. 
In the Hagg notation, a stacking fault is introduced simply 
through a change in sign. Deformation type stacking faults in 3C-SiC, 
4H-SiC and 6H-SiC systems, which are bordered by Shockley partials on the basal plane, are  (...$+++|-+++++$...), (...$++-- | -+--++--$...) 
and (...$+++---|-++---+++---$...), respectively, where the symbol 
`$|$' indicates the stacking fault plane(10). There are two 
types of (111)/(000l) planes where a deformation type stacking fault 
can form: (a) $A|a$ called as ``shuffle", and (b) $a|B$ called as 
``glide". 

Our calculations are based on first-principles
pseudopotentials within Density Functional Theory as 
implemented in the Plane Wave Self Consistent Field (PWSCF)\cite{Baroni} code. 
We employ a local density approximation (LDA) to exchange correlation energy
functional and use ultra-soft pseudopotentials\cite{USPP} 
to represent interaction 
between ionic cores and valence electrons, and a plane wave basis with an 
energy cut off of 30 Ry (180 Ry) in the representation of the Kohn-Sham 
wavefunctions 
(density). Supercells with 12 atomic planes for elemental semiconductors and 24
atomic planes for SiC are used in the calculations of configurations 
with of stacking 
faults. Corresponding Brillouin zones are rather small in the 
$z-$direction (perpendicular
to the plane of the fault) and integrals over them were sampled 
with a uniform 5$\times$5$\times$1 mesh. 
Positions of atoms inside the supercell were
relaxed to attain a minimum energy structure using Hellman-Feynman forces in the
Broyden, Fletcher, Goldfarb, Shanno (BFGS)-based method\cite{BFGS}. 
Structural parameters for Si, Ge and C in diamond structure 
and 3C, 4H and 6H polytypes of SiC agree with the experimental 
values within the typical LDA errors. We find the cohesive energies 
of all the SiC polytypes are rather close to each other
and both 6H and 4H structures are almost equal within 
the calculational errors and the lowest in energy.

We first determined generalized stacking fault 
energy ($\gamma-$) surfaces\cite{Vitek}
for the (111) or basal planes of all the crystals studied here. 
It corresponds to the energy required to displace a 
half of the crystal on one side of the given 
plane relative to the other and exhibits periodicity of the
crystal plane. For a given displacement path going from the origin to 
a burgers vector {\bf b}, a maximum in the $\gamma-$surface gives 
the the energy barrier for slip along that path. The lowest of 
these barriers among the family of various paths
joining origin and {\bf b} is called the unstable stacking fault 
energy\cite{Rice} $\gamma_{us}$.  
A local minimum of the $\gamma-$surface defines an 
intrinsic ( also called deformation(10) ) stacking
fault. We used fourier transform to analyze 
the $\gamma-$surface by sampling it on a uniform 
5$\times$5 mesh in the planar unit cell. 

\begin{table}[h]
\caption{Calculated and experimental estimates of intrinsic ($\gamma_{isf}$) and unstable ($\gamma_{us}$)
stacking fault energies for the slip in glide plane.}

\begin{center}
\begin{tabular}{ccccccccccccccccccc}
\hline
\hline
&& System &&       $\gamma_{isf}$(calc.)  && && && $\gamma_{isf}$(expt.)       && $\gamma_{us}$      && \\
&&        &&       (mJ/m$^2$)                            && && && (mJ/m$^2$)      && (J/m$^2$)    && \\ \hline

&& Si     &&        46.9                                  && && && 69\cite{Foll,Peter}                            &&1.7                          && \\ 

&& C      &&       250                                  && && && 279\cite{Pirouz,Peter}                         &&5.5                          && \\

&& Ge     &&        48.5                                  && && && ----                                           &&1.6                          && \\

&& 3C-SiC &&        10.1                                  && && && ----                                           &&2.8                          && \\

&& 4H-SiC &&         9.1                                  && && && 14.7$\pm$2.5\cite{Hong}                        &&2.9                          && \\

&& 6H-SiC &&         2.6                                  && && && 2.5$\pm$0.9\cite{Maeda}                        &&2.9                          && \\
\\ \hline \hline

\end{tabular}
\end{center}
\label{Dielectric-table}
\end{table}

In all the tetrahedral systems considered here, 
our results rule out the possible
slip on shuffle plane as it is energetically 25-36 times 
more expensive than the slip on the glide plane. 
The $\gamma-$surfaces of the glide plane of all 
tetrahedral semiconductors are topologically similar
(see Fig. 1 for the $\gamma-$surface of SiC). 
We find $\gamma_{us}$ is about the same for the three polytypes 
of SiC, which is about half the $\gamma_{us}$ of diamond and 
larger by 80 \% than that of Si and Ge (see Table I).
The trend in our estimates of $\gamma_{us}$ suggests that 
nucleation of dislocations on (111) planes is much easier 
in Si and Ge than in SiC and diamond\cite{Rice}.
The energy of intrinsic stacking fault located at 
(2/3,1/3) in the $\gamma-$surface ($\gamma_{isf}$), bears 
a rather different trend: it is smallest for the polytypes of SiC and 5 times 
larger for Si and even larger for diamond.
These values generally  agree well with experimental estimates 
and other calculations\cite{Kaxiras} wherever
available. Significantly lower values of $\gamma_{isf}$ (and somewhat large 
$\gamma_{us}$) of SiC support  large areas of
stacking faults which sometimes extend right upto the crystal boundaries with 
partials disappearing at the surface.

We estimated the free energies of perfect and faulted SiC with $ F =  E + F_{config}+F_{vib},$ where E is the total energy obtained with DFT calculations,
$F_{config}$ and $F_{vib}$ are the configurational and 
vibrational contributions to free energy respectively. 
\begin{figure}[h]
\includegraphics[scale=0.70]{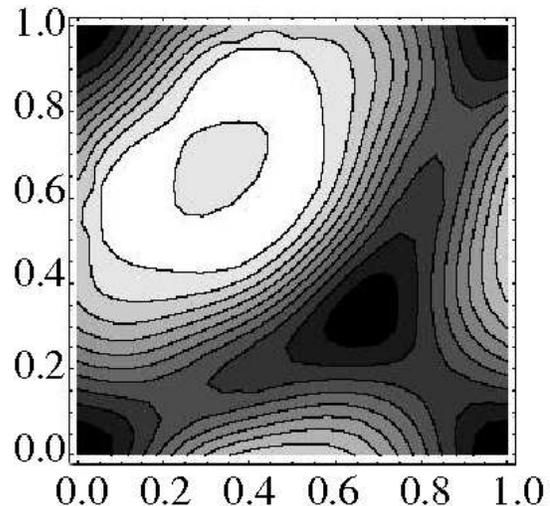}
\caption{Contour plot of Generalized Stacking Fault Energy($\gamma$) Surface for slip on the glide plane of 4H-SiC}
\end{figure}
$F_{config}$ is obtained from the detailed structure 
of basins of the $\gamma-$surface 
centered at  (0,0) and (2/3,1/3) corresponding to perfect 
and faulted structures.
We approximated the $\gamma-$surface in the neighbourhood of 
the centres ($x_0$, $y_0$) of these basins
(sampled by the system at finite temperature $T$) 
with a symmetry invariant parabolic form:$E(x,y) = 
E_{o} + K( (x-x_{o})^{2} + (y-y_{o})^{2}), $
to determine configurational contribution to free energy\cite{Lesar}
\begin{eqnarray}
F_{conf} = -k_{B} T Log( \frac{K}{2\pi k_{B}T} )
\end{eqnarray}
where $x$ and $y$ are the components of displacements for which 
the GSF has been obtained, the parameter $K$ was determined from 
DFT calculations carried out on a fine uniform $3\times3$ mesh over an area 
of 0.174 $bohr^{2}$ centered in each basin. Similarly, vibrational free 
energy is determined with
\begin{equation} 
F_{vib} = -k_{B} T \sum_{iq} Log(2sinh(\frac{\hbar\omega_{iq}}{2k_{b}T}))
\end{equation} 
$\omega_{iq}$ being the frequency of a phonon $i$ with wave 
vector $q$, obtained with DFT linear response 
calculations\cite{Baroni2}. We sampled $q$
at $\Gamma$ and $K$ points with weights of $\frac{1}{3}$ and 
$\frac{2}{3}$ respectively, omitting acoustic modes at the $\Gamma$.

The stacking fault energy as a function of temperature is obtained using:
$\gamma_{s}(T) = \gamma_{s}(0) + \Delta F_{conf}(T) + \Delta F_{vib}(T)$
where $\gamma_{s}(0)$ is the stacking fault energy at zero temperature, 
and $\Delta F$ is the
difference in free energy of the perfect and the faulted structures. 
In the approximation that phonon frequencies do not change 
with temperature, free energies
vary linearly with temperature at low T 
For 4H-SiC, we find the vibrational contribution to $\Delta F/T$   
(-0.2037 $\times 10^{-3}J/m^{2} K$) dominating over the 
configurational contribution (0.27$\times 10^{-5}J/m^{2} K$). 
Evidently, vibrational entropy plays a crucial role in stabilizing 
the faulted 4H-SiC structure at high 
temperatures. The stacking fault energy as a function of 
temperature (see Fig. 3) reveals that the faulted 
structure is energetically favorable at 60 K, 
supporting the observed stacking fault expansion and the
suggestion that it is driven by the free energy difference\cite{MKDas}. 

We determine the precise microscopic mechanism that stabilizes the faulted
structure by examining mode-by-mode contribution to the 
vibrational free energies.
In comparison with the perfect 4H structure, there 
emerge four soft phonon modes (see Figure 2) at 
$\Gamma$ point in the faulted 4H structure with frequencies
107.3, 107.4, 109.9, 110.0 $cm^{-1}$, which lowers its 
free energy through its contribution to the
entropy.
We note that most of these soft modes involve shear straining 
of the structure over a longer length-scale
in a direction perpendicular to the plane of the fault. 
These soft modes couple with shear strain and
can drastically reduce shear elastic moduli which 
in this case is $C_{44}$.  Thus, 
large shear strains can be induced even with
small shear stresses in the system. We note that our estimate 
of the temperature above which the faults get stabilized 
can have large errors due to many approximations made 
in our analysis. However, the mechanism we have 
identified here is expected to be reliable. In contrast 
to the 4H structure, we find the configurational 
contribution to free energy alone is adequate
in stabilizing faults in the 6H structure. This means the 
energy basin of the perfect 6H structure is stiffer than 
the energy basin of its faulted form yielding larger 
configurational entropy to the faulted form.

\begin{figure}[t]
\includegraphics[scale=0.33]{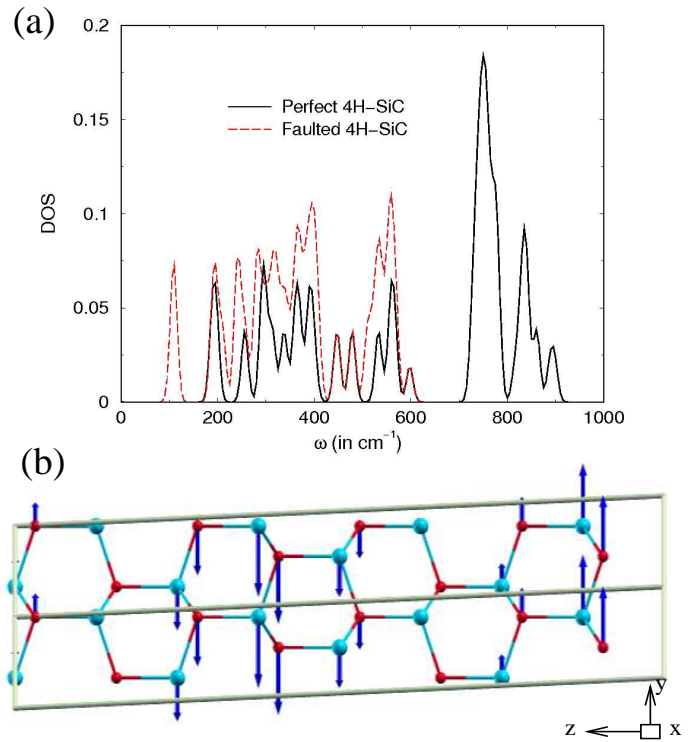}
\caption{(a) Phonon Density of States of perfect and faulted 4H-SiC (b) Soft 
shear phonon mode of frequency 107.3 $cm^{-1}$ in faulted 4H-SiC}
\label{Structure1}
\end{figure}

Finally, we would like comment on the relative 
stability of 4H and 6H structures of SiC.
Many groups\cite{Lambrecht,CHPark,P.KÃ¤ckell,CCheng1} 
have reported energies of these
structures within a couple of meV per atom of each other. 
 However, these are very small energies and we think both structures are 
equally stable at T=0 K within our
calculational errors. By including vibrational contribution 
to free energy at nonzero temperature, 
we find that the 6H structure is more stable than the 4H 
structure by a sizeable difference in free energy
(with a rate of 6$\times10^{-5}$ eV/atom/K). 
Along with our earlier conclusion that the 6H structure
 would also tend to form faults at finite temperature, 
we conclude that both 4H and 6H structures of SiC are metastable 
at finite temperature.
\begin{figure}[t]
\includegraphics[scale=0.33,angle=270]{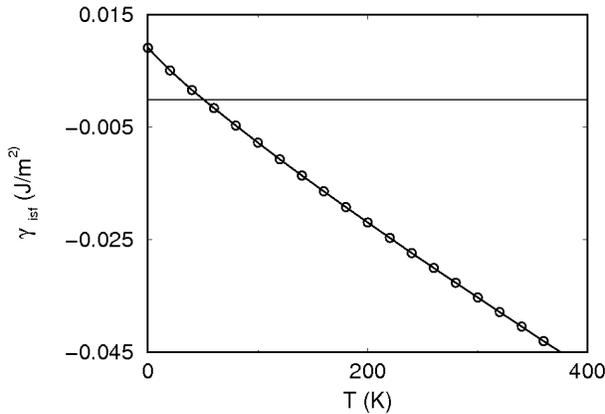}
\caption{Stacking Fault energy ($\gamma_{isf}$) of 4H-SiC as 
a function of temperature}
\label{Structure1}
\end{figure}

In summary, we have verified the speculation based on experimental 
work\cite{MKDas} that free energy difference is the driving force of stacking 
fault expansion in SiC. The stability of the faults has been shown to arise 
from the vibrational entropy and particularly soft phonon modes that involve
atomic displacements that shear the faulted structure on long length-scale.
As a result, stacking faults are expected to grow at finite temperature in
both 4H and 6H polytypes of SiC and this is a fundamental limitation of SiC 
for use in devices. Our work should stimulate further theoretical and 
experimental work that focus on destabilizing stacking faults in SiC through
suitable doping or other means.

Authors thank late Prof. F. R. N. Nabarro 
for stimulating discussions. TT acknowledges useful suggestions 
from Aarti. S, J. Bhattacharjee, N. Ellegard, 
M. Upadhyay Kahali and G. Dutta. We are grateful to
the CCMS and the central computer Lab at JNCASR for the 
use of computational facilities.

\end{document}